\documentclass[article,prl,twocolumn]{revtex4-1}
\usepackage{graphicx}
\usepackage{amsmath}
\usepackage{fancyhdr}
\usepackage{cancel}
\usepackage{color}
\setcitestyle{round}
\usepackage{ulem}
\begin{document}
\author{Purushottam D. Dixit}
\thanks{Corresponding author. Email: pd2447@columbia.edu}

\affiliation{Department of Systems Biology, Columbia University,\\ New York, NY 10032}

\author{ Abhinav Jain and Gerhard Stock}
\affiliation{Institute of Physics and \\  Freiburg Institute for Advanced Studies (FRIAS) \\ Albert Ludwigs University, \\ Freiburg, 79104 Germany}

\author{Ken A. Dill}
\affiliation{Laufer Center for Physical and Quantitative Biology,\\Department of Chemistry,\\ and Department of Physics and Astronomy,\\Stony Brook University,\\ Stony Brook, NY, 11790}

\title{Inferring transition rates on networks with incomplete knowledge}
\begin{abstract}
Across many fields, a problem of interest is to predict the transition rates between nodes of a network, given limited stationary state and dynamical information.   We give a solution using the principle of Maximum Caliber.  We find the transition rate matrix by maximizing the path entropy of a random walker on the network constrained to reproducing a stationary distribution and a few dynamical averages.  A main finding here is that when constrained only by the mean jump rate,  the rate matrix is given by a square-root dependence of the rate, $\omega_{ab} \propto \sqrt{p_b/p_a}$, on $p_a$ and $p_b$, the stationary state populations at nodes $a$ and $b$.  We give two examples of our approach.  First, we show that this method correctly predicts the correlated rates in a biochemical network of two genes, where we know the exact results from prior simulation.  Second, we show that it correctly predicts rates of peptide conformational transitions, when compared to molecular dynamics simulations.  This method can be used to infer large numbers of rates on known networks where smaller numbers of steady-state node populations are known.
\end{abstract}
\maketitle

\section{Introduction}
We are interested in an inference problem in network science.  Given the topology of a network and stationary populations at the nodes, what is the best model that can infer the rates of the dynamical flows along the edges?  Here are examples.  First, consider a spin  model with a known stationary distribution, for example, those used in neuroscience~\citep{schneidman2006weak}, protein evolution~\citep{shekhar2013spin}, or colloidal sciences~\citep{han2008geometric}. It is of great interest to  infer the best dynamical process that is consistent with a given rate of spin flip. Second, in systems biology, we often know the topology of a network of metabolites, or proteins or regulatory elements.  In addition, ``-omics'' experiments can estimate the abundances of the many metabolites or proteins or regulatory elements at the nodes during the steady-state functioning of a cell.  However, this information alone is not sufficient to explain cell function.  We also need to know the forward and backward rates of fluxes $\omega_{ab}$ and $\omega_{ba}$ between all nodes $a$ and $b$, for example in metabolic networks~\citep{orth2010flux}.    Measuring all these rates is  practically impossible at present, particularly for large networks.  Third, in structural biology, it is common to perform computer simulations of the conformations of biomolecules and infer Markov models among metastable states from those simulations~\citep{chodera2014markov}.  Here, computing the populations of the states can be done rapidly, whereas computing the kinetic barriers between them is much slower.  

In many such problems, the popular approach, especially for large networks, is to hypothesize a parametric dynamical model and learn the parameters of this model from data.  Often, a large amount of data is required and the parameters learnt are not unique.  We treat this problem as a matter of inference, in the spirit of statistical mechanics, where full distribution functions are inferred from a few measured equilibrium-averaged properties~\citep{Press2012,peterson2013maximum}.   We provide a solution employing  the dynamical analog of the principle of Maximum Entropy, Maximum Caliber, seeking the single best model that is consistent with under-determined data.

Mathematically, we seek a Markovian random walker on a network that has the maximal path entropy and that otherwise satisfies  prescribed constraints.  Towards that goal, we first define a class of walkers that satisfy a) a prescribed stationary state distribution $\{ p_a \}$ over the nodes $\{ a \}$ of a network and b) certain dynamical properties defined over the ensemble of stationary state paths $\{ \Gamma \}$. Examples of dynamical properties include the average distance travelled by the walker per unit time,  the average number of reactions per per unit time, or the average number of amino acid changes per unit time in a constantly evolving protein.

Below, we first derive the Maximum Caliber Markov process. We then illustrate its predictive power with two examples, a gene expression network and a network of  metastable states of a small peptide.

\section{Theory}
Consider a Markovian random walker on a directed network $G$ with nodes $V = \{ a \}$ and edges $E$. Assume that the random walker has a unique stationary state distribution $\{ p_a \}$ over nodes $\{ a\}$ that is independent of the initial conditions. The instantaneous probability of the walker being at a node $b$ at time $t$, $q_b(t)$,  is governed by
\begin{eqnarray}
\frac{dq_b(t)}{dt} &=& \sum_{a} \omega_{ab} q_a(t) - \sum_{a} \omega_{ba} q_b(t) =  \sum_a {\bf \Omega}_{ba} q_a(t) \label{eq:mas0}
\end{eqnarray}
where the time independent rate of transition $\omega_{ab}$ from node $a$ to node $b$ is non zero if and only $(a, b) \in E$.

In order to maximize the path entropy, we discretize time into time intervals $\delta t$ and at a later time, take the limit $\delta t \rightarrow 0$.  In the discrete time scenario, the transition rates $\omega_{ab}$ with units of inverse time are replaced by unitless transition probabilities  $k_{ab}$. The matrix ${\bf k}$ of transition probabilities depends on the discretization interval and is given by ${\bf k} = e^{{\bf \Omega} \delta t} \approx  {\bf I} + {\bf \Omega} \delta t.$ The second approximation is accurate only for  $\delta t \ll 1$. Here, ${\bf I}$ is the identity matrix.

Let us define an ensemble $\{ \Gamma \}$ of stationary state paths $\Gamma \equiv \dots \rightarrow a \rightarrow b \rightarrow c \rightarrow d \dots$ of the walker of total duration $T$. The entropy of this ensemble is $\mathcal S_T =  -\sum_{\Gamma} P(\Gamma) \log P(\Gamma) =  -T \sum p_a k_{ab} \log k_{ab}$~\citep{filyukov1967method,dixit2014inferring,cover2012elements}. The time normalized path entropy $\mathcal S$ is
\begin{eqnarray}
\mathcal S &=& -\sum_{(a,b) \in E} p_a k_{ab} \log k_{ab}. \label{eq:s0}
\end{eqnarray}
The sum in Eq.~\ref{eq:s0} is only taken over edges $(a,b) \in E$ of the network. All summations below are also restricted to edges $(a,b) \in E$ unless otherwise stated.

\subsection{The constraints on the paths}
Any discrete time Markov process with $\{ p_a \}$ as the stationary distribution and  $k_{ab}$ as transition probabilities satisfies two sets of linear constraints (normalization and stationarity). These constraints are understood as follows: First, from state $i$ at time $t$, the system {\it has to} land at one of the states $j$ at time $t+\delta t$. Second, at stationary state, a system in state $j$ at time $t+\delta t$ comes from one of the state $i$. We have,
\begin{eqnarray}
\sum_{b} k_{ab} =1~\forall~a~{\rm and}~\sum_a p_a k_{ab} = p_b~\forall~b.\label{eq:c0}
\end{eqnarray}
Another important constraint is detailed balance, $p_a k_{ab} = p_b k_{ba}$. Below, we will see how detailed balanced constrained can be applied explicitly~\citep{dixit2014inferring}. 

We introduce a node-connectivity variable $N$ such that $N_{aa} = 0$, $N_{ab} = 1$ {\it if} $(a,b) \in E$.  The path ensemble average of $N$ over all trajectories $\Gamma$ is given by~\citep{dixit2014inferring,filyukov1967method}
\begin{eqnarray}
\langle N \rangle &=& \sum_{(a,b) \in E} p_a k_{ab} N_{ab}. \label{eq:th}
\end{eqnarray}
$\langle N \rangle$ is the mean number of transitions in a single time step $\delta t$.  Below we use $\langle N \rangle$ to take the desired limit $\delta t \rightarrow 0$ in order to convert the discrete time Markov chain to a continuous time Markov process.

Also, we constrain the ensemble average $\langle r^{i}_{ab} \rangle$ of  arbitrary dynamical rate variables $r^{i}_{ab}$. Examples of dynamical constraints include the average distance travelled by a particle diffusing on an energy landscape per unit time, the number of spin flips per unit time in a spin glass model, the  average number of reactions per unit time, etc. We assume $r^{i}_{aa} = 0$ and $r^{i}_{ab} = 0$ if $(a,b) \notin E$. $r^{i}_{aa} = 0$ is a mere convenience and not a strict requirement of our development (see supplementary materials for details).  Given that the network is directed, in general we have $N_{ab} \neq N_{ba}$ and $r^{i}_{ab} \neq r^{i}_{ba}$.

\subsection{Maximizing the path entropy, subject to constraints.} We now maximize the path entropy $\mathcal S$ (Eq.~\ref{eq:s0}) with respect to transition
 probabilities $k_{ab}$ subject to constraints imposed by Eq.~\ref{eq:c0} and Eq.~\ref{eq:th}. Using the method of Lagrange multipliers, we write the unconstrained Caliber $\mathcal C$~\citep{stock2008maximum,Press2012}
\begin{eqnarray}
\mathcal C &=& \mathcal S + \sum_a m_a \left (\sum_b p_a k_{ab} - p_a \right ) +  \sum_b l_b \left (\sum_a p_a k_{ab} - p_b \right ) \nonumber \\ &-& \gamma
 \left ( \sum_{a,b} p_a k_{ab} N_{ab} - \langle N \rangle \right )-\sum_i \rho_i \left ( \sum_{a,b} p_a k_{ab} r^{i}_{ab} - \langle r^{i}_{ab} \rangle \right ). \nonumber \\ \label{eq:c1}\end{eqnarray}

We maximize the Caliber with respect to $k_{ab}$ and derive a closed form expression for the transition probabilities  (see supplementary materials for a detailed derivation)
\begin{eqnarray}
k_{ab} &=& \mu \delta t \sqrt{\frac{f_b g_a}{f_a g_b}} \sqrt{\frac{p_b}{p_a}} {\bf \Delta}_{ab} ~{\it if}~ (a,b) \in E~{\rm and} 
\end{eqnarray}
and $k_{aa} = 1-\sum_b k_{ab}$. Here $\mu  \delta t= e^{-\gamma}$ and ${\bf \Delta}_{ab} = e^{ -\sum_i \rho_i r^{i}_{ab}}$~{\it if} $(a,b) \in E$ and zero otherwise. Constants $f_a$ and $g_a$ are determined from self-consistent equations
\begin{eqnarray}f_a &=& \sum_b {\bf \Delta}_{ab} \sqrt{\frac{f_bp_b}{g_b}}~{\rm and}~g_a = \sum_b {\bf \Delta}_{ba} \sqrt{\frac{g_bp_b}{f_b}} \label{eq:fg_selfc}
\end{eqnarray}

Since ${\bf k} = {\bf I} + {\bf \Omega} \delta t$ as $\delta t \rightarrow 0$. We take the limit and get
\begin{eqnarray}
\omega_{ab} &=&  \mu \sqrt{\frac{f_b g_a}{f_a g_b}} \sqrt{\frac{p_b}{p_a}}{\bf \Delta}_{ab}~{\it if}~ (a,b) \in E  \label{eq:omega}
\end{eqnarray}
and $\omega_{aa} = -\sum_b \omega_{ab}$. Eqs.~\ref{eq:omega} is the most general result of this work. Since constants $f_a$ and $g_a$ are determined from {\it global} self-consistent equations Eq.~\ref{eq:fg_selfc}, the transition rate $\omega_{ab}$ between any two states $a$ and $b$ depends on the structure of the entire network.

\subsection{Detailed balance}

Detailed balance constraint requires $p_a \omega_{ab} = p_b \omega_{ba}$. Thus, if both $(a,b)$ and $(b,a)$ are not in $E$, $\omega_{ba} = \omega_{ab} = 0$ and the connection between $(a,b)$ can be removed. Consequently, we assume that the network $G$ is undirected i.e. $(a,b) \in E \Rightarrow (b,a) \in E$ and vice versa. Imposing detailed balance constraints on transition probabilities $k_{ab}$ is equivalent to constraining symmetrized forms of the dynamical variables $r^{i\dag}_{ab} = \frac{1}{2} \left (r^i_{ab} + r^i_{ba} \right ) $ (see ~\citep{dixit2014inferring} for a proof). In this case, we have  ${\bf \Delta} = {\bf \Delta}^{\rm T}$ and $\bar f = \bar g$. Thus
\begin{eqnarray}
\omega_{ab} =  \mu  \sqrt{\frac{p_b}{p_a}}{\bf \Delta}_{ab}~{\it if}~(a,b) \in E. \label{eq:main2}
\end{eqnarray}
It is easy to check that Eq.~\ref{eq:main2} satisfies detailed balance. Interestingly,  the transition rates of a detailed balanced Markov process are determined entirely locally from the properties of states $a$ and $b$ alone and does not depend on the global structure of the network. In contrast, the same transition rate depends on the structure of the entire network when detailed balance is not satisfied (see Eq.~\ref{eq:omega}).

When we constrain only the mean jump rate (${\bf \Delta}_{ab} = 1~{\it if}~(a,b)\in E$ and zero otherwise), the transition rates are given by the simple expression:
\begin{eqnarray}
\omega_{ab} =  \mu  \sqrt{\frac{p_b}{p_a}}~{\it if}~ (a,b) \in E.
\end{eqnarray}
This result should be contrasted with other popular functional forms of the transition rates that satisfy a prescribed stationary distribution, e.g. Glauber dynamics~\citep{glauber1963time} and Metropolis dynamics~\citep{mariz1990comparative}. For example, historically,  Glauber developed his dynamics to study magnetic spins~\citep{glauber1963time}.  The particular form of the transition rates was motivated by ``a desire for simplicity''~\citep{glauber1963time}. The derivation of square root dynamics presented here, while not resorting to a {\it ad hoc} prescription for transition rates, follows the same intuitive principle i.e. finding the simplest model that is consistent with a given set stationary and dynamical constraints and surprisingly predicts a dynamics that is {\it qualitatively} different from  Glauber and Metropolis dynamics.

\subsection{Application to theories of chemical reaction rates}

Our predicted square-root dynamics has an interesting interpretation in theories of chemical reaction rates.  In organic chemistry, so-called {\it extra-thermodynamic relationships} express empirical observations about how the rates and mechanisms of certain types of chemical reactions are related to their equilibria.  These go by the name of the Bronsted relation (often applied to acid-base catalysis)~\citep{leffler2013rates}, the Polanyi relationship for surface catalysis~\citep{michaelides2003identification}, Marcus theory for electron transfer reactions in solution~\citep{marcus1968theoretical}, or $\Phi$-value analysis for protein folding~\citep{matouschek1993application}.  In short, these approaches express that rates are related to equilibrium constants in the form of
\begin{eqnarray}
 \omega (x) = c K (x)^{\alpha}  \label{eq:bronsted}
\end{eqnarray}
where $x$ is a variable representing a systematic change, such as a series of different acids of different $p K_a$'s, $K(x)$ is the equilibrium constants for that series, and $\omega(x)$ are the rates of reaction. $\alpha$ is a value that usually ranges from 0 to 1, expressing the degree of resemblance of the transition state to the products of the reaction.

Now consider a two state system where state $a$ are reactants and state $b$ are products such that $p_b > p_a$. We have neglected the population in the transition state since it is much smaller than the state populations at equilibrium. We have equilibrium constant $K = p_b / p_a > 1$. If no dynamical constraints are imposed,  the transition rate $\omega_{ab}$ of the $a\rightarrow b$ reaction according to the MaxCal process is given by $\omega_{ab} = \mu \sqrt{\frac{p_b}{p_a}} = \mu K^{1/2}$. In short, our MaxCal approach predicts that $\alpha = 1/2$ in Eq.~\ref{eq:bronsted} is the most parsimonious assumption for the reaction mechanism prior to any knowledge of global rate information, implying that the transition state is halfway between reactants and products.

\section{Illustrating Eqs.~\ref{eq:omega} and ~\ref{eq:main2} with two examples.}

We test the model predictions on a problem of correlated expression of two genes that are transcribed by the same promoter and on a problem of the equilibrium conformations of a small peptide.

\subsection{Expression dynamics of two correlated genes}

\begin{figure}
\includegraphics[scale=0.15]{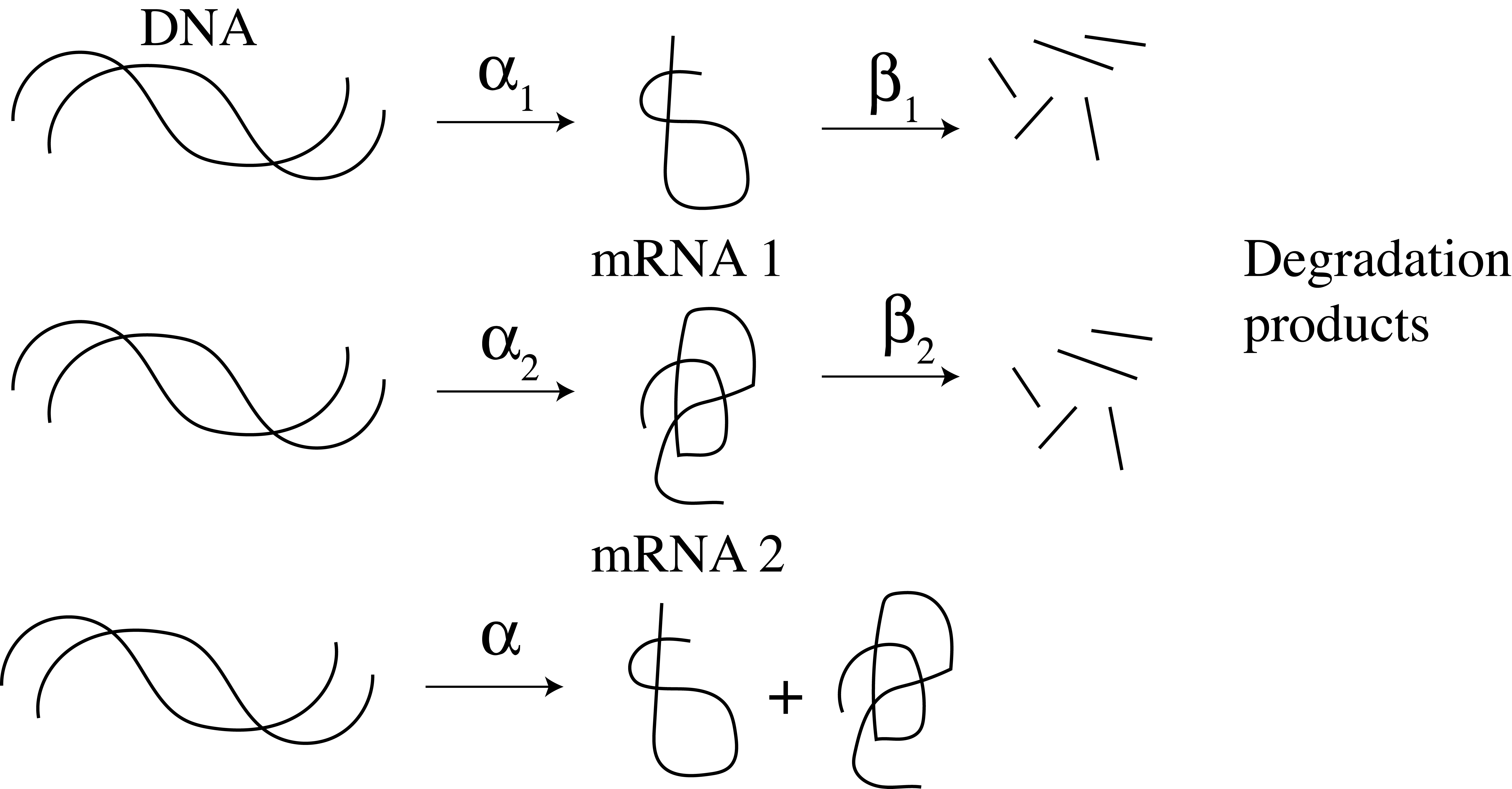}
\caption{Schematic of two genes that are simultaneously transcribed by the same promoter. Rates $\alpha_1$, $\alpha_2$ ,and $\alpha$ represent the synthesis rates and rates $\beta_1$ and $\beta_2$ represent the degradation rates.\label{fg:promoter}}
\end{figure}

It is difficult to infer the underlying regulatory architecture of large biochemical networks from stationary populations of the components  such as mRNAs and proteins.  Here, we show how Eq.~\ref{eq:omega} can accurately predict the full chemical master equation (CME) from stationary distributions and overall rate parameters.  Consider a biochemical circuit where two genes are adjacent to a constitutively expressing promoter region. We assume that the genes are either transcribed individually or simultaneously and that they are degraded individually.  In this toy example, two stochastic variables are correlated, namely the copy numbers of the two mRNA molecules. We first construct a CME to mimic the biochemical circuit {\it in silico} (Fig.~\ref{fg:promoter}). There are 5 rate parameters in the CME: 3 synthesis rates $\alpha_1$, $\alpha_2$, and $\alpha$ and two degradation rates $\beta_1$ and $\beta_2$.  If $n_1$ and $n_2$ are the number of molecules of the first and the second mRNA,  the chemical master equation describing the system has the following form
\begin{eqnarray}
\frac{dp(n_1,n_2;t)}{dt} &=& \alpha_0 \left ( p(n_1-1,n_2) - p(n_1+1,n_2) \right ) \nonumber \\ &+& \alpha_1 \left ( p(n_1,n_2-1) - p(n_1,n_2+1) \right ) \nonumber \\ &+&\alpha \left (p(n_1-1,n_2-1) - p(n_1,n_2) \right ) \nonumber \\ &+& \beta_1 \left ( (n_1 + 1)p(n_1+1,n_2) - n_1 p(n_1,n_2) \right ) \nonumber \\ &+& \beta_2 \left ( (n_2 + 1)p(n_1,n_2+1) - n_2 p(n_1,n_2) \right ) \nonumber \\ \label{eq:cme}
\end{eqnarray}
Here, $p(n_1, n_2;t)$ is the instantaneous probability of having $n_1$ and $n_2$ molecules of mRNA 1 and 2 respectively at time $t$. The terms correspond to individual synthesis of mRNA 1 and mRNA 2, the simultaneous synthesis of both mRNAs, and the degradation of mRNA 1 and mRNA 2.

We assume that we can experimentally estimate the joint probability distribution $p_{ss}(n_1,n_2)$ of the mRNA copy numbers at steady state. Additionally, we also assume that we have two reporters that count the total number of expression events and the total number of degradation events respectively. Note that the reporters are agnostic to which of the two RNAs has been synthesized or degraded. With only these three pieces of information, can we estimate the transition rate matrix of the system?

\begin{figure*}
        {\bf (A)}
        \includegraphics[scale=0.4]{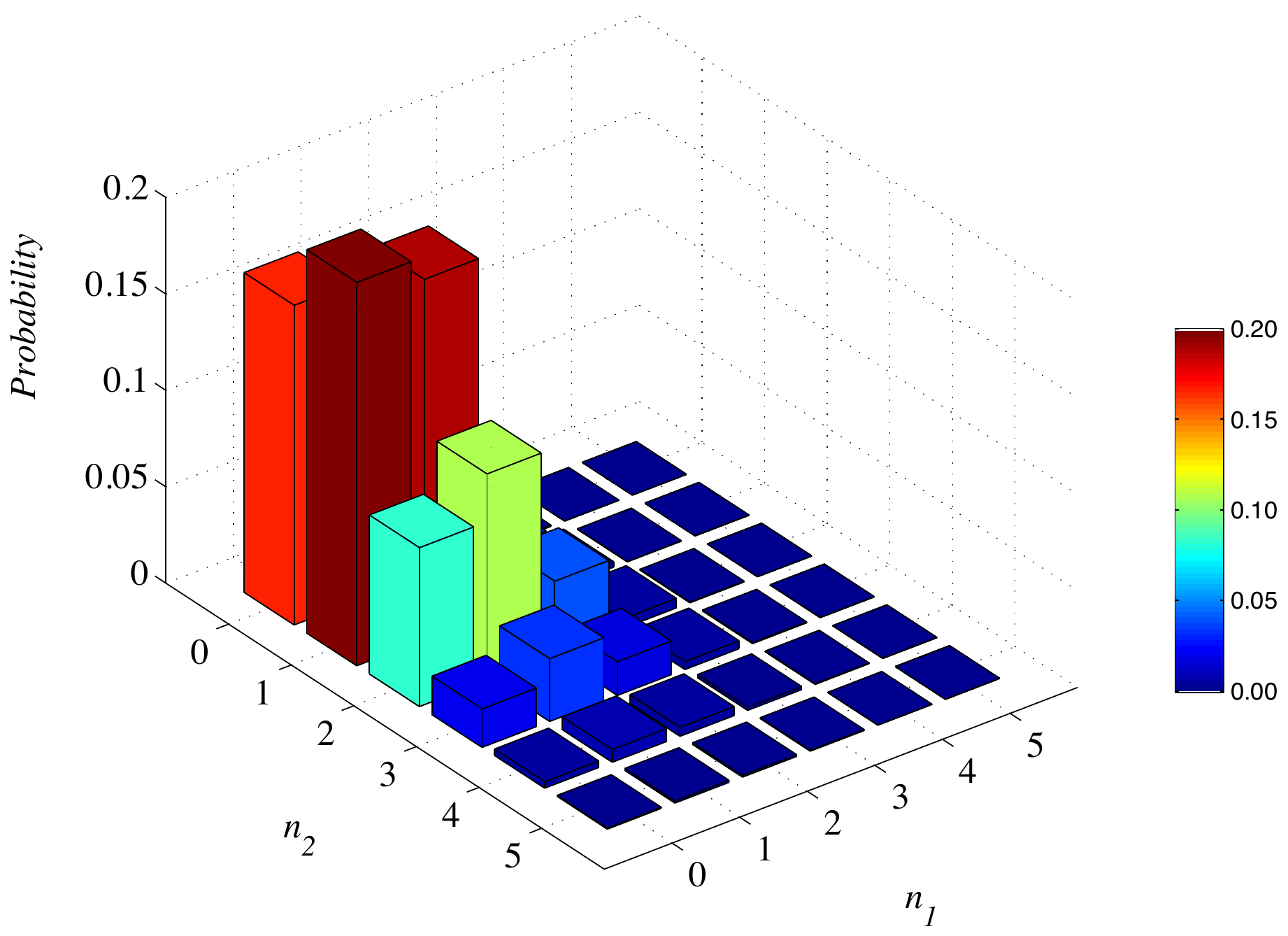}\hspace{5mm}
        {\bf (B)}
        \includegraphics[scale=0.45]{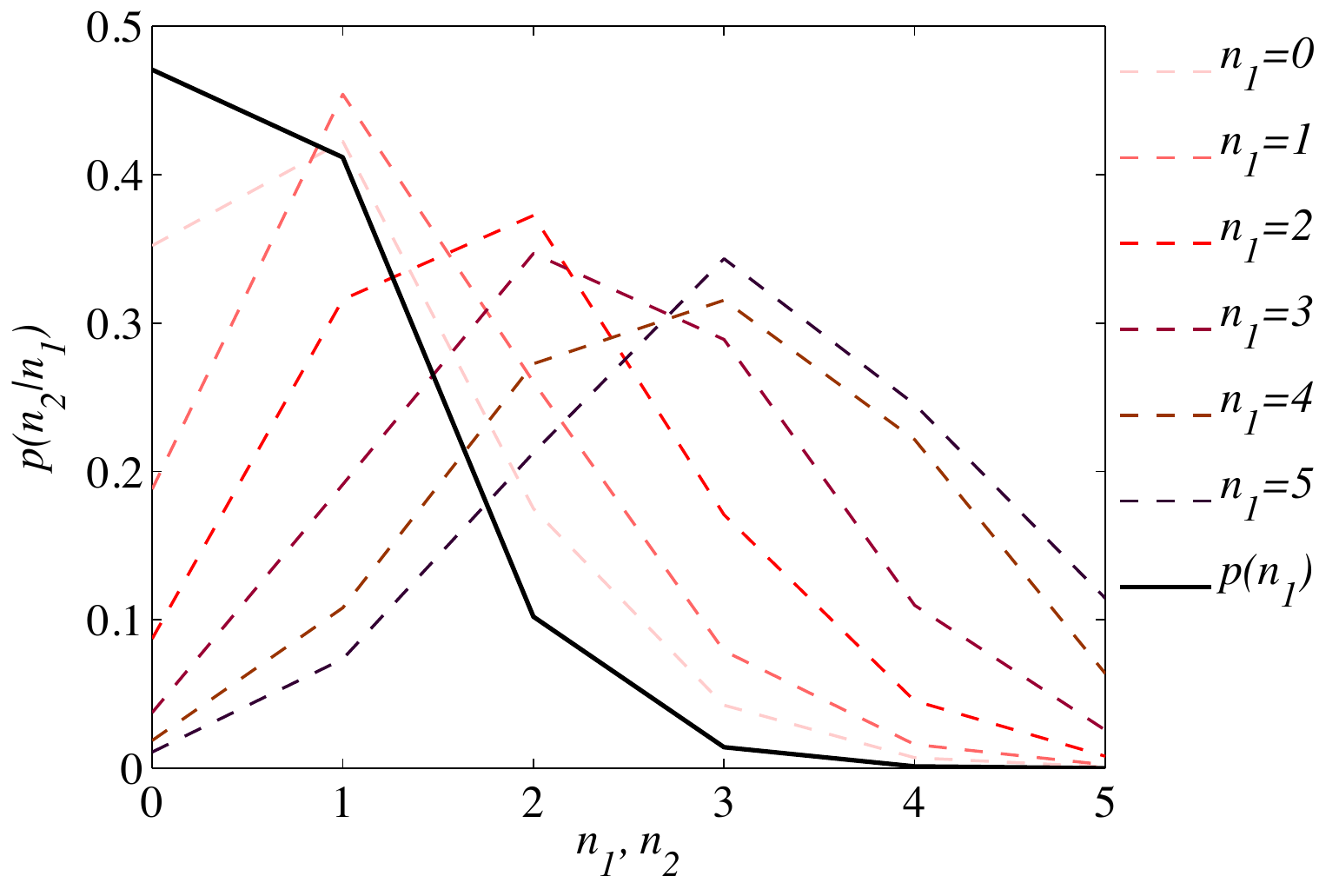}
        \caption{Panel A: The joint stationary state distribution $p_{ss}(n_1,n_2)$ when the parameters are set at $\left ( \alpha_1, \alpha_2 , \alpha , \beta_1 , \beta_2 \right ) = (1, 0.5, 2.5, 5, 10)$. Panel B: The conditional probability $p(n_2|n_1)$ at different values of $n_1$.  \label{fg:joint_prob}}
\end{figure*}

We choose the following parameters for the CME:  $\left ( \alpha_1,\alpha_2 , \alpha , \beta_1 , \beta_2 \right ) = (1, 0.5, 2.5, 5, 10)$. The choice ensures that the number of any of the two mRNA molecules is limited to $< 6$. This results in  a small system size in this proof of principle work where the total number of states is $6\times 6 = 36$. In Fig.~\ref{fg:joint_prob} panels A and B, we show the numerically observed joint distribution $p_{ss}(n_1,n_2)$. The correlated pattern of expression  is apparent in panel B; the probability $p_{ss}(n_2|n_1)$ depends on $n_1$. The higher the value of $n_1$, the higher $n_2$ values become more probable as a result of the correlated expression.

In the CME, while there are only 5 rate parameters, there are 145 possible transitions. There are 30 + 30 transitions that correspond to synthesis of mRNA 1 or 2. There are 30 + 30 transitions that correspond to degradation of mRNA 1 or 2 and there are 25 transitions that correspond to simultaneous synthesis of both mRNAs.  Each transition has its own rate constant but not all rate constants are independent of each other. For example, the transition $(2, 3) \rightarrow (2,2)$ has a rate constant $3\beta_2$ and the transition $(2,4)\rightarrow (2,3)$ has a rate constant $4\beta_2$. Moreover, many other transition rates are equal to each other, for example, the rates for transitions that increase the first mRNA copy number  such as $(1,2) \rightarrow (2,2)$ and $(3,1) \rightarrow (4,1)$ are equal to $\alpha_1$ and so on.
\begin{figure}
        \includegraphics[scale=0.45]{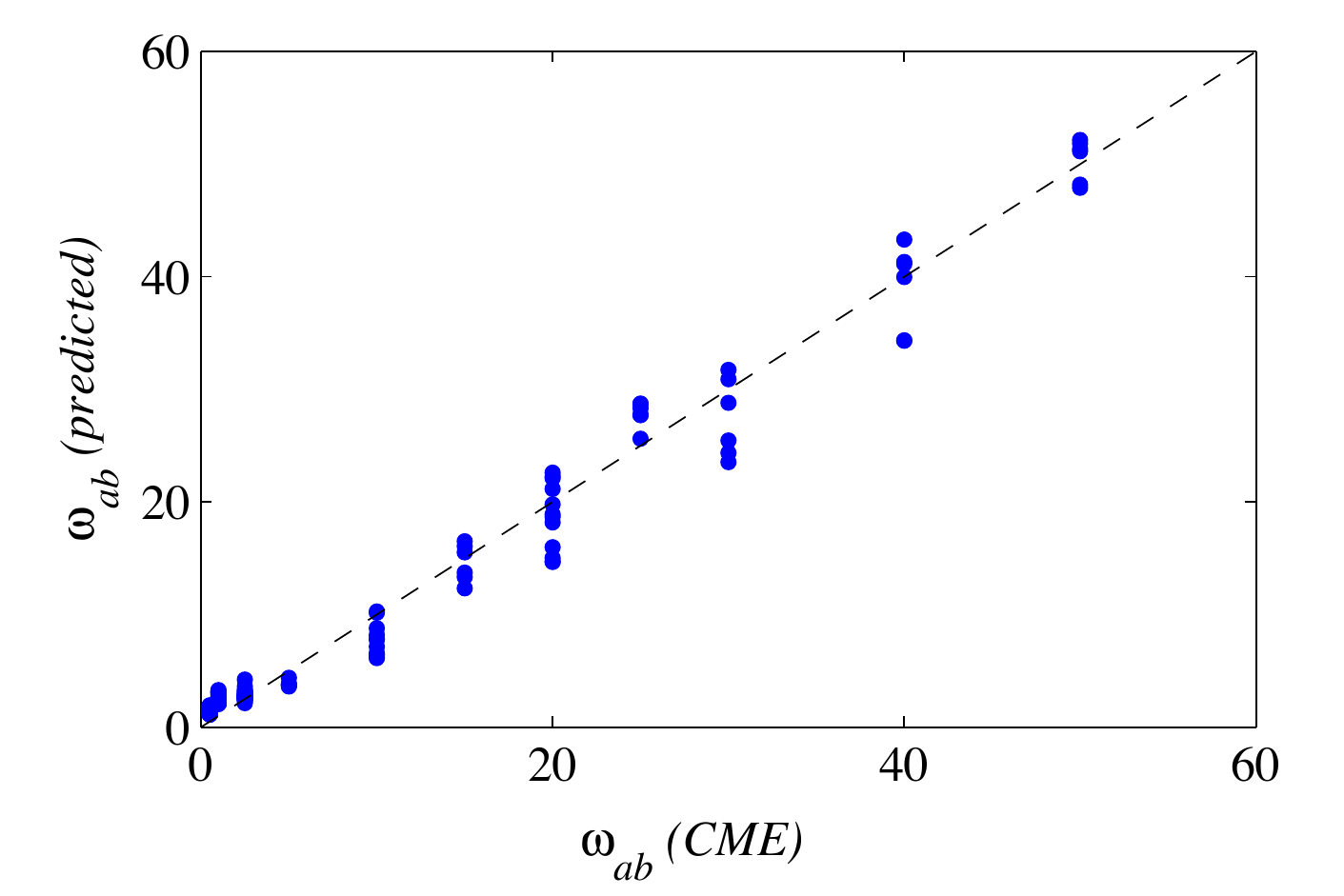}
        \caption{Predicted rates vs. toy model values for all 145 possible transitions in the gene expression model. \label{fg:pred_CME}}
\end{figure}

We find that Eq.~\ref{eq:omega} accurately gives the full 145 rate parameters, without the knowledge of the differential rates of synthesis and degradation of the two mRNAs (see supplementary materials for details of the fitting procedure) (see Fig.~\ref{fg:pred_CME}).  Taken to larger scale, it implies that using steady state data on cell-to-cell variability in gene expression and a few overall kinetic measurements, we can a full set of chemical master equations and infer regulatory details.  This infered model is optimal in the Maximum Caliber sense.

\subsection{Dynamics of a small peptide} Second, we study the equilibrium dynamics of metastable states of a small peptide comprising 7 alanine amino-acid residues. A metastable state is an ensemble of geometrically and dynamically proximal microstructures that have a significant net population.  Classifying protein structures into their metastable states is an active area of research~\citep{lane2011markov,prinz2011markov,chodera2014markov,jain2012identifying}.

\begin{figure}
        \includegraphics[scale=0.4]{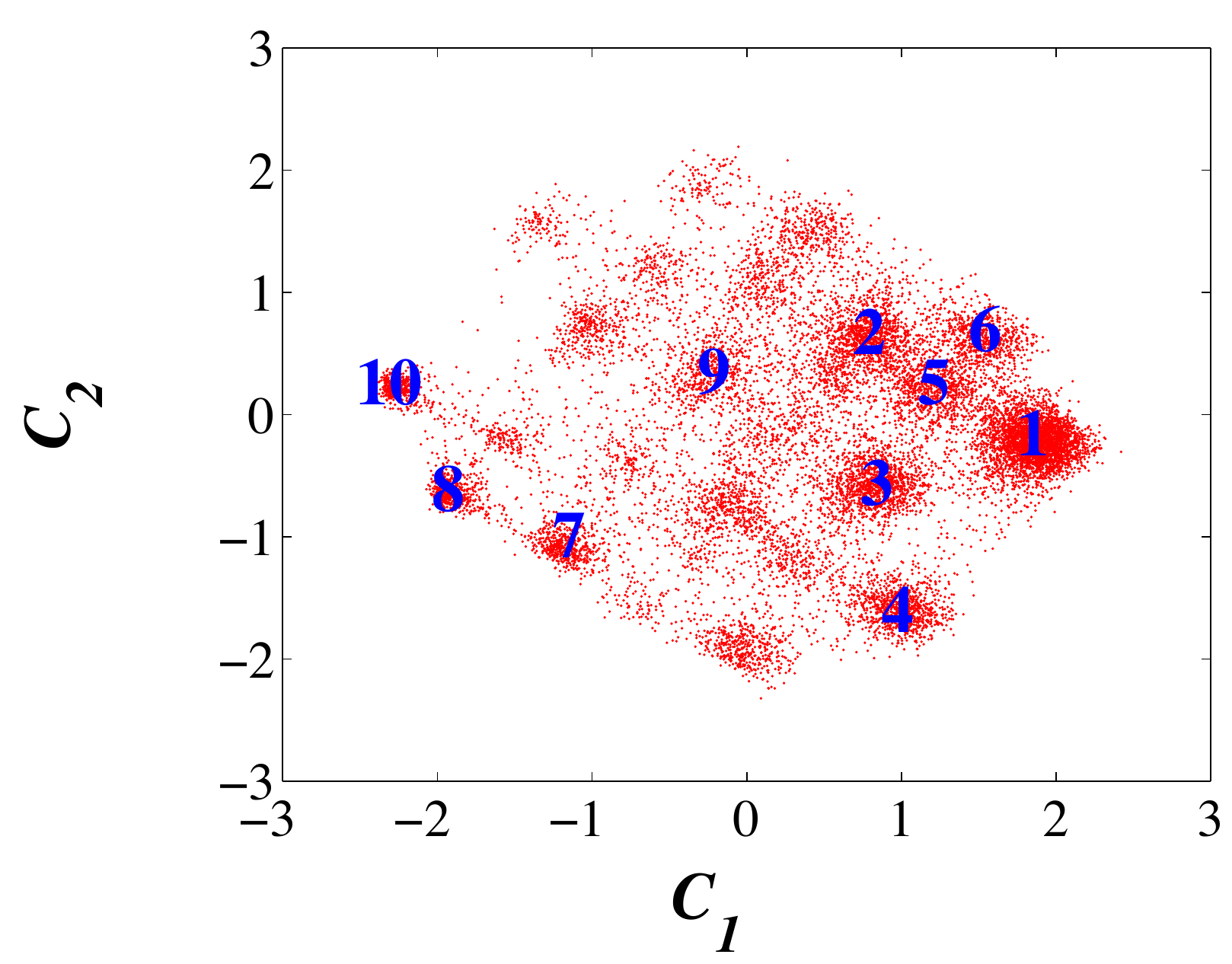}
        \caption{Two principal coordinates from molecular dynamics simulations of the alanine-7 peptide~\citep{altis2008construction}, showing that the microscopic structures can be lumped into well separated metastable states. The 10 most populated states are labelled in decreasing order of state probability. \label{fg:simulate}}
\end{figure}

As a test, we compare to previous extensive MD simulations of this peptide~\citep{altis2008construction}, which led to the identification of 32 metastable conformations~\citep{jain2012identifying} (see Fig.~\ref{fg:simulate}). From that MD simulation, we took the states as defined by Jain and Stock and estimated the relative probability $p_a$ of each metastable state $a$ as the fraction of time points in the full trajectory when the peptide was in that state. We also estimated the transition probabilities $k_{ab}$ as the fraction of the events in the trajectory when the peptide was in state $a$ at time $t$ and transitioned into state $b$ at time $t+\delta t$ ($\delta t = 1$ ps in a MD simulation of total duration $T=800$ ns).

To predict the transition probabilities $k_{ab}$, we first need to estimate the transition rate matrix ${\bf \Omega}$ using Eq.~\ref{eq:main2}. In order to guess the functional form of the transition rate matrix, we need to identify a dynamical constraint. The accuracy of our predictions depends on how well the dynamical constraint captures the diffusion of the peptide on the free energy landscape. Unfortunately, there is no systematic procedure to guess a `good' constraint variable. This is a common aspect of Maximum-Entropy methods. See~\citep{caticha2004maximum,caticha2012entropic,dixit2013quantifying} for a discussion on the role of constraints in maximum entropy methods.

To guess the dynamical constraint, we make two observations: the transition rate decreases when 1) when the average conformational separation between states is increased, keeping the stationary probabilities and the free energy barrier constant and 2) when the free energy barrier is increased, keeping the average conformational distance and the stationary state probabilities constant. As a first guess, we only model the effect of geometric separation and neglect the free energy barrier. We choose a simple geometric constraint $r_{ab}$: for any two metastable states $a$ and $b$ and microstates $x$ and $y$ such that $x \in a$ and $y \in b$, $r_{ab} = \left (\int_{x\in a, y\in b} ||x-y||_2 dx dy \right ) ^{\frac{1}{2}}$.   $r_{ab}$ is the mean distance between a randomly chosen microstate point $x$ in metastable state $a$ and randomly chosen microstate point $y$ in metastable state $b$. Similarly, $r_{aa}$ is the average distance between any two microstates $x$ and $y$ within a macrostate $a$; it represents the `volume' of  macrostate $a$. The distance between any two microstates $x$ and $y$ is defined as the Euclidean distance between their internal coordinate representation~\citep{jain2012identifying}.  The dynamical average  $\langle r_{ab} \rangle$  represents the average distance travelled by the microstates of the peptide per unit time.

\begin{figure*}
        \includegraphics[scale=0.5]{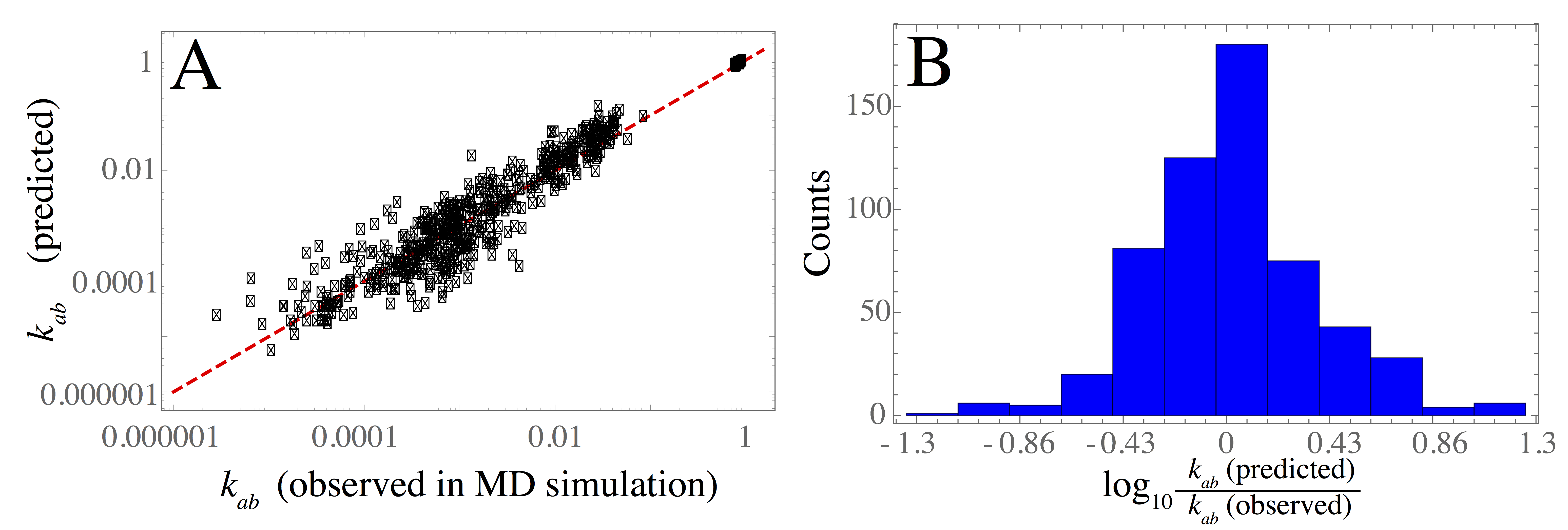}
        \caption{Panel A: A comparison of transition probabilities $\{ k_{ab} \}$ predicted using Eq.~\ref{eq:main2} and those observed in the MD simulation for the 7 residue peptide shows good agreement over 5 orders of magnitude. Panel B: Histogram of the logarithm of predicted vs observed transition probabilities shows that while the majority of the transition probabilities are predicted very accurately, virtually all transition probabilities are predicted within one order of magnitude. \label{fg:pred}}
\end{figure*}

Using Eq.~\ref{eq:main2}, the numerically estimated stationary probabilities, and the chosen geometric constraint, we arrive at the functional form of the transition rate matrix ${\bf \Omega}$. The transition matrix had two free parameters, $\mu$ the time scale and $\rho$ the Lagrange multiplier associated with $\langle r_{ab} \rangle$.  In order to estimate the transition probability matrix ${\bf k}$ from the predicted matrix of transition rates ${\bf \Omega}$, we need the discretization time scale $\delta t$, a third  parameter. Since $\mu$ and $\delta t$ can be combined together, we only needed to determine 2 parameters, $\rho$ and $\mu \delta t$. Note that the Maximum Caliber approach guesses the parametric form of the transition probabilities, one can use any suitable numerical technique and experimental information to estimate the parameters. Since we have access to microscopic data, in this proof of principle study, we used the entire transition probability matrix to fit the two free  parameters. Using multiple simulated annealing runs to minimize the total error between known and predicted transition probabilities, we found that the best fits were at $\mu \delta t = 45 \pm 5 $ and $\rho = 4.8 \pm 0.6$.  Fig.~\ref{fg:pred} shows that the rates predicted by the Max Cal approach quite accurately capture the correct values obtained from the full MD simulation, over 5 orders of magnitudes of rates.

\section{Discussion}
Here, we describe theory that takes a given network, steady-state populations on its nodes, and a couple of global dynamical constraints, and finds the microscopic transition rates among all the nodes.  We do this using Maximum Caliber, a Maximum-Entropy-like principle for dynamical systems.  A main finding is that the MaxCal transition rates are proportional to the square root of ratios of the state populations.  We illustrate our results on a toy gene expression network and peptide conformations, for which we know the correct rates in advance.  We believe this treatment could be useful in many areas of network modeling, including in spin-glass models of the immune system~\citep{shekhar2013spin,mora2010maximum}, colloidal assemblies~\citep{han2008geometric}, neuronal networks~\citep{schneidman2006weak}, master-equation models of noisy gene expression~\citep{paulsson2004summing,dixit2013quantifying}, and in the browsing behavior of web crawlers on the internet~\citep{baldi2003modeling}.

\section{Supplementary materials}

\subsection{Deriving the maximum Caliber transition rates}

For notational simplicity, we derive Eq.~\ref{eq:omega} for one dynamical constraints $r_{ab}$. Differentiating the Caliber in Eq.~\ref{eq:c1} with respect to $k_{ab}$ and setting the derivative to zero,
\begin{eqnarray}
p_a (\log k_{ab} + 1) &=& m_a p_a + l_b p_a - \gamma N_{ab} p_a - \rho r_{ab} p_a \nonumber \\ &\Rightarrow& k_{ab} =  \frac{\beta_a}{p_a} \lambda_b {\bf W}
_{ab} \label{eq:kijform}
\end{eqnarray}where $ \frac{\beta_a}{p_a} = e^{m_a - 1}$, $\lambda_b = e^{l_b}$, and $ {\bf W}_{ab} = e^{-\gamma N_{ab} - \rho r_{ab}}$.

For a given value of $\gamma$ and $\rho$, the Lagrange multipliers $\beta$s and $\lambda$s are determined by self consistently solving Eqs.~\ref{eq:c0}. We 
have\begin{eqnarray}\sum_b k_{ab} &=& 1 \Rightarrow \sum_b  {\bf W}_{ab} \lambda_b = \frac{p_a}{\beta_a}~~~\nonumber \\ {\rm and}~~ \sum_a p_a k_{ab} &=& p_
b \Rightarrow \sum_b  {\bf W}_{ba} \beta_b = \frac{p_a}{\lambda_a}. \label{eq:sc}
\end{eqnarray}
To further simplify Eqs.~\ref{eq:sc}, let usidentify $\bar \lambda$ and $\bar \beta$ as the vectors of Lagrange multipliers and define $\mathcal D[\bar x]_a
 = \frac{p_a}{x_a}$, a non-linear operator on vectors $
\bar x$. We have
\begin{eqnarray}
{\bf W} \bar \lambda &=& \mathcal D[ \bar \beta]~{\rm and}~{\bf W^{\rm T}} \bar \beta = \mathcal D[ \bar \lambda] \nonumber \\
\Rightarrow \mathcal D[{\bf W} \bar \lambda] &=& \bar \beta~{\rm and}~\mathcal D[{\bf W^{\rm T}} \bar \beta] = \bar \lambda \label{eq:sc1}
\end{eqnarray}

Now, we write ${\bf W} = {\bf I} + \mu \delta t {\bf \Delta}$ where $e^{-\gamma} = \mu  \delta t$, ${\bf \Delta}_{ab} = 0$ if $(a,b) \notin E$, and ${\bf \Delta}_{ab} = e^{-\rho r_{ab}}$ when $(a,b) \in E$. We have
\begin{eqnarray}
\frac{p_a}{\lambda_a + \mu \delta t f_a} = \beta_a~{\rm and}~\frac{p_a}{\beta_a + \mu \delta t g_a} = \lambda_a \label{eq:self1}
\end{eqnarray}
where
\begin{eqnarray}\bar f = {\bf \Delta} \bar \lambda~{\rm and}~ \bar g = {\bf \Delta}^{\rm T} \bar \beta. \label{eq:fg}
\end{eqnarray}
We solve these two equations algebraicly recognizing that $f_a$ and $g_a$ do not directly depend on $\lambda_a$ and $\beta_a$. The positive roots are
\begin{eqnarray}
\lambda_a &=& \frac{\sqrt{f_a} \sqrt{g_a} \sqrt{{\delta t}^2 f_a g_a \mu ^2+4 p_a}-{\delta t} f_a g_a \mu }{2 g_a} \\
\beta_a &=& \frac{\sqrt{f_a} \sqrt{g_a} \sqrt{{\delta t}^2 f_a g_a \mu ^2+4 p_a}-{\delta t} f_a g_a \mu }{2 f_a}.
\end{eqnarray}
Given that we're only interested in transition probabilities $k_{ab}$ up to orders of $\delta t$ and $k_{ab} = \mu \delta t \frac{\beta_a}{p_a} \lambda_b \Delta_{ab}$ already has a $\delta t$ term, we only need to the zeroth order terms for $\bar \lambda$ and $\bar \beta$ as $\delta t\rightarrow 0$. We solve the algebraic equations~\ref{eq:self1} simultaneously for $\lambda_a$ and $\beta_a$ in terms of $f_a$, $g_a$, $p_a$, and $dt$. We then take only the zeroth order terms in $\delta t$ (see supplementary materials for details). We have
\begin{eqnarray}
\lambda_a = \sqrt{\frac{f_ap_a}{g_a}} ~{\rm and}~\beta_a = \sqrt{\frac{g_ap_a}{f_a}} \label{eq:lagrangem2}
\end{eqnarray}
Eq.~\ref{eq:lagrangem2} and Eq.~\ref{eq:fg} can be self-consistently solved for $\bar \lambda$ and $\bar \beta$ and we can obtain $\bar f = {\bf \Delta}\bar \lambda$ and $\bar g = {\bf \Delta^{\rm T}}\bar \beta$.

\subsection{When dynamical constraints are such that $r_{aa} \neq 0$}

As mentioned in the main text, we have assumed for convenience that the dynamical constraints $r_{ab}$ is such that $r_{aa} = 0$. Here, we show that when $r_{aa} \neq 0$, the maximum entropy problem is equivalent to constraining a modified constraints $r^{\dag}_{ab} =  r_{ab} - \frac{1}{2} \left ( r_{aa} + r_{bb} \right) $.

We start with recognizing ${\bf W} = {\bf J} + \mu \delta t {\bf \Delta}$ as above where ${\bf J}$ is a diagonal matrix such that ${\bf J}_{aa} = e^{-\rho r_{aa}}$. We have
\begin{eqnarray}
\frac{p_a}{j_a\lambda_a + \mu \delta t f_a} = \beta_a~{\rm and}~\frac{p_a}{j_a \beta_a + \mu \delta t g_a} = \lambda_a
\end{eqnarray}
where $j_a = {\bf J}_{aa}$ and $\bar f = {\bf \Delta} \bar \lambda~{\rm and}~ \bar g = {\bf \Delta}^{\rm T} \bar \beta$. Again solving to zeroth order in $\delta t$, we get
\begin{eqnarray}
\lambda_a = \sqrt{\frac{f_a p_a}{g_a j_a}}~{\rm and}~\beta_a = \sqrt{\frac{p_a g_a}{f_a j_a}}.
\end{eqnarray}
Finally, the transition probability $k_{ab}$ is given by
\begin{eqnarray}
k_{ab} &=& \mu \delta t \frac{\beta_a}{p_a} \lambda_b  e^{-\rho r_{ab}} =  \mu \delta t \sqrt{\frac{p_b}{p_a}} \sqrt{\frac{ f_b g_a}{f_a g_b }} \frac{1}{\sqrt{j_a j_b}}   e^{-\rho r_{ab}  }   \nonumber \\&=& \mu \delta t \sqrt{\frac{p_b}{p_a}} \sqrt{\frac{ f_b g_a}{f_a g_b }}    e^{-\left (\rho r_{ab}  - \frac{1}{2} \left ( r_{aa} + r_{bb} \right ) \right )} \nonumber \\ 
&=&\mu \delta t \sqrt{\frac{p_b}{p_a}} \sqrt{\frac{ f_b g_a}{f_a g_b }}  e^{-\rho r^{\dag}_{ab}} \label{eq:omega2}
\end{eqnarray}
where  $r^{\dag}_{ab} =  r_{ab} - \frac{1}{2} \left ( r_{aa} + r_{bb} \right) $. Comparing Eq.~\ref{eq:omega2} to Eq.~\ref{eq:omega}, it is clear that the problem of constraining a dynamical constraint $r_{ab}$ such that $r_{aa}$ is non-zero is equivalent to constraining a modified dynamical constraint $r^{\dag}_{ab} =  r_{ab} - \frac{1}{2} \left ( r_{aa} + r_{bb} \right) $. Note that by definition, $r^{\dag}_{aa} = 0$. Thus, for convenience, we assume that this transformation is already performed and $r_{aa} = 0$.

\subsection{Constructing and fitting the maximum entropy Markov proces for the gene network}

From the numerical experiments, we obtain accurate estimate of the stationary state probability distribution $p_{ss}(n_1,n_2)$. Let $a \equiv (n_1, n_2) $ and $b \equiv (n_1^\dag, n_2^\dag)$ be any two states of the system. We know that there is a directed edge from state $a$ to state $b$ {\it iff}
\begin{enumerate}
       \item Synthesis or degradation of mRNA 1: $n_1 = n_1^\dag \pm 1$ and $n_2 = n_2$
       \item Synthesis or degradation of mRNA 2: $n_1 = n_1^\dag$ and $n_2 = n_2^\dag \pm 1$
       \item Simultaneous synthesis: $n_1^\dag = n_1 + 1$ and $n_2^\dag = n_2 + 1$
\end{enumerate}
As mentioned in the main text, the `experiments' tell us the total number of degradation and synthesis events per unit time but do not have the ability to distinguish between the two mRNAs. We constrain two quantities, the number of degradation events per unit time and the number of synthesis events per unit time. Accordignly, we set ${\bf \Delta}$ in Eq.~\ref{eq:omega} as follows
\begin{enumerate}
       \item No edge between nodes $a$ and $b$: ${\bf \Delta}_{ab} = 0~{\rm if}~(a, b) \not \in E $
       \item Synthesis events: ${\bf \Delta}_{ab} = \eta~{\rm if}~n_1 = n_1^\dag~{\rm and}~n_2+1 = n_2^\dag~{\rm or}~n_1+1 = n_1^\dag~{\rm and}~n_2 = n_2^\dag~{\rm or}~~n_1+1 = n_1^\dag~{\rm and}~n_2+1 = n_2^\dag$
       \item Degradation of mRNA 1: ${\bf \Delta}_{ab} = n_1 \zeta ~{\rm if}~n_1 = n_1^\dag+1~{\rm and}~n_2 = n_2^\dag$
       \item Degradation of mRNA2: ${\bf \Delta}_{ab} = n_2 \zeta ~{\rm if}~n_1 = n_1^\dag~{\rm and}~n_2 = n_2^\dag +1$
\end{enumerate}
Here, $\eta \ge 0$ and $\zeta \ge 0$ are exponentials of Lagrange multipliers similar to those used in the main text. The Lagrange rate constant $\mu$ in Eq.~\ref{eq:omega} is assumed to be absorbed in the Lagrange multipliers. The next step is to determine $\bar f$, $\bar g$, $\bar \lambda$, and $\bar \beta$ using numerically estimated $p_{ss}(n_1,n_2)$ and ${\bf \Delta}$. In order to determine the transition rate matrix ${\bf \Omega}$ for any value of $\eta$ and $\zeta$, we solve Eq.~\ref{eq:fg} and Eq.~\ref{eq:lagrangem2} self consistently.

Given that we have access to all rate constants in this proof of principles work, we minimize the error between the known rate constants of Eq.~\ref{eq:cme} and those predicted by Eq.~\ref{eq:omega} by varying $\eta$ and $\zeta$ using a simulated annealing protocol. We find that $\eta \approx 0.4$ and $\zeta \approx 26.5$ resulted in the best agreement between the known and the predicted rate constants. Multiple simulated annealing runs predicted rates that were identical to the ones shown in the main text. Note that although we used the entire transition matrix to learn the Lagrange multipliers, it is equally possible to learn them from global dynamical constraints.

\end{document}